\documentclass[preprint,aps,amsmath]{revtex4-1}
\usepackage{graphicx}

\begin{document}

\title{The Moran model as a dynamical process
on networks and its implications for neutral speciation}

\author{Marcus A.M. de Aguiar$^{1,2}$ and Yaneer Bar-Yam$^1$}

\affiliation{$^1$New England Complex Systems Institute, Cambridge,
Massachusetts 02142 \\$^2$ Instituto de F\'{\i}sica `Gleb Wataghin',
Universidade Estadual de Campinas, Unicamp\\ 13083-859, Campinas, SP,
Brasil}

\begin{abstract}

In genetics the Moran model describes the neutral evolution of a
bi-allelic gene in a population of haploid individuals subjected to
mutations. We show in this paper that this model can be mapped into an
influence dynamical process on networks subjected to external influences.
The panmictic case considered by Moran corresponds to fully connected
networks and can be completely solved in terms of hypergeometric
functions. Other types of networks correspond to structured populations,
for which approximate solutions are also available. This new approach to
the classic Moran model leads to a relation between regular networks based
on spatial grids and the mechanism of isolation by distance. We
discuss the consequences of this connection for topopatric speciation
and the theory of neutral speciation and biodiversity. We show that the
effect of mutations in structured populations, where individuals can mate
only with neighbors, is greatly enhanced with respect to the panmictic
case. If mating is further constrained by genetic proximity between
individuals, a balance of opposing tendencies take place: increasing
diversity promoted by enhanced effective mutations versus decreasing
diversity promoted by similarity between mates. Stabilization occurs with
speciation via pattern formation. We derive an explicit relation
involving the parameters characterizing the population that indicates
when speciation is possible.

\end{abstract}

\maketitle

%%%%%%%%%%%%%%%%%%%%%%%%%%%%%%%%%%%%%%%%%%%%%%%%%%%%%%%%%
%%%%%%%%%%%%%%%%%%%%%%%%%%%%%%%%%%%%%%%%%%%%%%%%%%%%%%%%%
\section{Introduction}
\label{intro}

A basic problem in population genetics is to predict how allele
frequencies change in a population according to the underlying rules
governing reproduction. For very large populations the Hardy-Weinberg
law applies and no change is expected between consecutive generations.
However, for finite populations this is not necessarily true, and drift
can play an important role.

One of the first models to describe genetic drift in a finite
population is the Wright-Fisher model \cite{ewens}. It considers a
population of $N$ diploid individuals and a single gene with two
alleles $A_0$ and $A_1$, so that there are a total of $2N$ genes. Given
that the number of alleles $A_1$ in the population at time $t$ is $i$,
one can easily compute the probability to have $j$ alleles $A_1$ at
time $t+1$. Assuming that reproduction occurs by randomly picking $2N$
genes among the previous population with replacement and that there is
no mutation, this probability is given by the binomial distribution
\begin{displaymath}
p_{ij} = \left( \begin{array}{c} 2N \\
j \end{array} \right) (i/2N)^j[1-(i/2N)]^{2N-j}.
\end{displaymath}

These {\it transition probabilities} form a matrix whose eigenvalues
and eigenvectors contain all the information about the evolution of the
system. Although the Wright-Fisher matrix is rather complicated,
several analytical results can be extracted from it and even mutations
can be included \cite{ewens}.

Other models were developed later that allowed for simpler mathematical
treatment than the Wright-Fisher model or its generalization by
Cannings \cite{cann74}. Of particular importance is the Moran model
\cite{moran,ewens,jw}, which considers haploid individuals and
overlapping generations. Here a single hermaphroditic individual
reproduces at each time step, with the offspring replacing the expiring
parent. The transition probabilities can also be written down
explicitly and all its eigenvalues and eigenvectors can be calculated
for the case of zero mutations \cite{wat61,glad78}. When mutations are
included the eigenvalues of the transition matrix and the stationary
probability distribution, corresponding to the first eigenvector, can
still be calculated \cite{cann74,gill}.

Here we show that the Moran model can be mapped into a dynamical
problem on networks, putting this classic model of population genetics
in a broader and modern perspective. The mapping takes a panmictic
population into a fully connected network, where the dynamical problem
can be completely solved in terms of generating functions
\cite{aguiar2005,aguiar2007}. This provides a simple and elegant
representation of the complete set of eigenvectors of the problem. The
connection with the network dynamics gives, to our knowledge, the first
complete solution of the Moran model.

Networks that are not fully connected map into non-random mating in
structured populations. In particular, regular networks based on
two-dimensional grids relate to spatially structured populations where
mating is allowed only between neighbors. This, in turn, provides the
basic mechanism of isolation by distance, as first proposed by Sewall
Wright \cite{wright1943}. It has been recently shown \cite{aguiar2009}
that this process can lead to speciation, termed topopatric speciation, and
that the patterns of diversity that arise are fully compatible with the
characteristics of biodiversity observed across many types of
species in nature \cite{rosenzweig1995} .
Although no exact solution exists for the Moran model for structured
populations, approximate solutions do exist for the equivalent network
problem \cite{aguiar2007}. In this paper we explore this connection to
discuss the mechanisms underlying  topopatric speciation \cite{aguiar2009}.

The paper is organized as follows: in sections \ref{network} and
\ref{master} we define the network dynamical system associated to the
Moran process and write down its master equation and transition
probabilities. In section \ref{map} we show how the Moran model can be
mapped into this network problem. In section \ref{equi} we summarize the
Moran-network properties: the distribution of allele frequencies at
equilibrium, with its mean value and variance, and the limit of
large populations. In section \ref{structured} we discuss approximations
for other network topologies and, in section \ref{speciation}, their
consequences for speciation.

%%%%%%%%%%%%%%%%%%%%%%%%%%%%%%%%%%%%%%%%%%%%%%%%%%%%%%%%%
%%%%%%%%%%%%%%%%%%%%%%%%%%%%%%%%%%%%%%%%%%%%%%%%%%%%%%%%%
\section{The network dynamical system}
\label{network}

Networks are mathematical structures composed of nodes and links
between the nodes. The nodes often represent parts of a system and the
links the interaction between the parts. Networks can model a wide
range of systems in biology, engineering and the social sciences
\cite{bararev}. In this work we will associate nodes to a particular
gene carried by individuals in a population and links will be
established between individuals that can mate with each other. In this
section networks will be treated as mathematical abstractions with a
particular dynamics of network states; the connection with
population genetics will be established in section \ref{map}, although the
correspondence with the Moran process is going to become evident as we
proceed.

Consider a network with $N+N_0+N_1$ nodes. To each node $i$ we assign
an internal state $x_i$ which can take only the values $0$ or $1$. The
nodes are divided into three categories: $N$ nodes are free to change
their internal state (according to the rule stated below); $N_1$ nodes
are frozen in the state $x_i=1$ and $N_0$ nodes are frozen in $x_i=0$.
The frozen nodes are assumed to be connected to all free nodes and we
consider them as perturbations to the `free' network, composed of the
free nodes only. The information about the free network topology is
contained in its adjacency matrix ${\cal A}$ defined as ${\cal
A}_{ij}=1$ if nodes $i$ and $j$ are connected, ${\cal A}_{ij}=0$ if
they are not and ${\cal A}_{ii}=0$. We refer to the free nodes
connected to $i$ as their neighbors. The degree $k_i = \sum_j {\cal
A}_{ij}$ is the number of neighbors of node $i$.

The dynamics on the free nodes is defined as follows: at each time step
a node is selected at random to be updated. With probability $p$ the
state of the node does not change, and with probability $1-p$ it copies
the state of one of its connected nodes, selected randomly among the
$k_i$ free neighbors or $N_0+N_1$ frozen nodes. If the node to be
updated is $i$, then
\begin{displaymath}
x_i^{t+1} = \left\{
\begin{array}{l}
x_i^t \qquad \mbox{with probability      } p \\
x_j^t \qquad \mbox{with probability      } \frac{1-p}{k_i+N_0+N_1}
\end{array} \right.
\end{displaymath}
where $j$ is connected to $i$.

We call this process an {\it influence dynamics}, since the state of a
node changes according to the state of its neighbors. This system can
model a number of interesting situations, such as, for example:\\

(a) An election with two candidates where part of the voters have a
fixed opinion while the others change their intention according to the
opinion of the others.\\

(b) A sexually reproducing population of $N$ haploid individuals where
the internal state represents two alleles of a gene. Taking $p=1/2$,
the update of a node mimics the mating of the focal individual with one
of its neighbors. The focal individual is replaced by the offspring,
which can take the allele of each parent with $50\%$ probability. Since
the free node can also copy the state of a frozen node, the values of
$N_0$ and $N_1$ can be associated with mutation rates, as we will show
later.\\

(c) A ferromagnetic material composed of atoms with magnetic moment
$\pm 1/2$ interacting with an external magnetic field. \\

Although the influence process is very simple, its analysis can be quite
complicated for networks of arbitrary topology. We will first consider
the simpler case of fully connected networks, where ${\cal A}_{ij}=1$
if $i \neq j$, ${\cal A}_{ii}=0$ and $k_i=N-1$. Later we will discuss
the consequences of other topologies and provide approximate results
for these cases using the fully connected case as a basis.

%%%%%%%%%%%%%%%%%%%%%%%%%%%%%%%%%%%%%%%%%%%%%%%%%%%%%%%%%
%%%%%%%%%%%%%%%%%%%%%%%%%%%%%%%%%%%%%%%%%%%%%%%%%%%%%%%%%
\section{Master equation and transition probabilities}
\label{master}

For fully connected networks the nodes are indistinguishable and there
are only $N+1$ global states, that we call $\sigma_k$, $k=0,1,...,N$.
The state $\sigma_k$ has $k$ free nodes in the state $1$ and $N-k$ free
nodes in the state $0$. There is no need to count the frozen nodes,
since they never change. If $P_t(m)$ is the probability of finding the
network in the state $\sigma_m$ at the time $t$ then, $P_{t+1}(m)$ can
depend only on $P_t(m)$, $P_t(m+1)$ and $P_t(m-1)$, since only one node
is updated per time step. According to the updating rule above, the
dynamic of the probabilities is described by the following equation:
\begin{displaymath}
\begin{array}{ll}
P_{t+1}(m) &=  \displaystyle{P_t(m)\left\{p + \frac{(1-p)}
{N(N+N_0+N_1-1)} \left[ m(m+N_1-1) +
(N-m)(N+N_0-m-1) \right] \right\} +} \\ \\
& P_t(m-1) \displaystyle{\frac{(1-p)}{N(N+N_0+N_1-1)}
(m+N_1-1)(N-m+1)} \, + \\ \\
& P_t(m+1) \displaystyle{\frac{(1-p)}{N(N+N_0+N_1-1)}
(m+1)(N+N_0-m-1)} \;.
\end{array}
\end{displaymath}

The term inside the first brackets gives the probability that the state
$\sigma_m$ does not change in that time step and is divided into two
contributions: the probability $p$ that the node does not change plus
the probability $1-p$ that the node does change. In latter case, the
state of the node is $x_i=1$ with probability $m/N$, and it may copy a
different node in the same state, $x_j=1$, with probability
$(m-1+N_1)/(N+N_0+N_1-1)$. Also, if $x_i=0$, which has probability
$(N-m)/N$, it may copy another node $x_j=0$ with
probability $(N-m-1+N_0)/(N+N_0+N_1-1)$. The other terms are obtained
similarly.

The probabilities $P_t(m)$ define a $P_t$ vector of $N+1$ components.
In terms of $P_t$ the above master equation can be written in matrix
form as
\begin{displaymath}
P_{t+1} = U P_t \equiv \left[ 1 - \frac{(1-p)}{N(N+N_0+N_1-1)} A\right]
P_t
\end{displaymath}
where the {\it evolution matrix } $U$, and also the auxiliary matrix
$A$, is tri-diagonal. The non-zero elements of $A$ are independent of
$p$ and are given by
\begin{displaymath}
\begin{array}{l}
A_{m,m} = 2m(N-m) + N_1(N-m) + N_0 m \\
A_{m,m+1} = -(m+1)(N+N_0-m-1) \\
A_{m,m-1} = -(N-m+1)(N_1+m-1).
\end{array}
\end{displaymath}
These transition elements are the analogue of the Wright-Fisher
transition probabilities described in the Introduction for the network
dynamics.

Let $\vec{a}_r$ and $\vec{b}_r$ be the right and left eigenvectors of
$U$ (and therefore of $A$) and $\lambda_r$ the corresponding
eigenvalues, so that $U \vec{a}_r = \lambda_r \vec{a}_r$ and $U^T
\vec{b}_r= \lambda_r \vec{b}_r$. The transition probability between two
states $\sigma_M$ and $\sigma_L$ after the time $t$ can be written as
\begin{equation}
P(L,t;M,0) = \sum_{r=0}^N  b_{rM} a_{rL}
\lambda_r^t \;. \label{problm}
\end{equation}
where $a_{rL}$ and $b_{rM}$ are the components of the right and left
r-th eigenvectors. The eigenvalues of $U$ are given by
\begin{displaymath}
\lambda_r = 1 - \frac{(1-p)}{N(N+N_0+N_1-1)}\mu_r
\end{displaymath}
where $\mu_r$ are the eigenvalues of $A$. Equation (\ref{problm})
indicates that the $\lambda_r$ have to be smaller or equal to 1,
otherwise $P(L,t;M,0)$ would eventually become larger than 1. Moreover,
the eigenvectors corresponding to $\lambda=1$ completely determine the
asymptotic behavior of the system, since the contributions of all the
others to $P(L,t;M,0)$ die out at large times.

The eigenvalues of $A$ are given by \cite{aguiar2007}
\begin{displaymath}
\mu_r = r(r-1+N_0+N_1)\;,
\end{displaymath}
which indeed implies that $0 \leq p \leq \lambda_r \leq 1$. Therefore,
if and only if $N_0=N_1=0$ there are two asymptotic (absorbing) states,
corresponding to $r=0$ and $r=1$, given by $\sigma_0$ (all node in
state 0) and $\sigma_N$ (all nodes in state 1). Otherwise there is only
one possible asymptotic state, corresponding to $r=0$. All
other eigenvectors, related to the transient dynamics, can be
calculated explicitly in terms of hypergeometric generating functions
\cite{aguiar2007}. We do not write them down here because we are only
interested in equilibrium properties.

%%%%%%%%%%%%%%%%%%%%%%%%%%%%%%%%%%%%%%%%%%%%%%%%%%%%%%%%%%%%%%%%%%%%%
%%%%%%%%%%%%%%%%%%%%%%%%%%%%%%%%%%%%%%%%%%%%%%%%%%%%%%%%%%%%%%%%%%%%%
\section{Mapping the Moran model onto network dynamics}
\label{map}

In order to map the evolution of a panmictic population of $N$
hermaphroditic individuals into the fully connected network problem
described above we use the following notation: we associate $x_i$ to
the allele of the haploid individual $i$, which is either 0 for allele
$A_0$ or 1 for allele $A_1$. At each time step a random individual $i$
is chosen to reproduce, and a random mate $j$ is selected among the
remaining $N-1$ individuals. The focal individual $i$ is then replaced
by the offspring.

Reproduction is carried out in two steps. The first step is the sexual
reproduction itself: with probability $1/2$ the allele $x_i$ is passed
to the offspring and with probability $1/2$ it takes the value $x_j$.
The second step takes mutation into account: after having taken the
allele of the focal individual or its mate, the allele might change,
from 0 to 1 with probability $\mu_-$ or from 1 to 0 with probability
$\mu_+$. This corresponds to the Moran model with asymmetric mutations
and is very similar to the influence process previously described for
networks. In the framework of networks, the update of the node by keeping
its own state or copying the state of a free neighbor corresponds to
sexual reproduction. Copying the state of a frozen node represents
mutation and depends on $N_0$ and $N_1$.

However, the two processes are not quite the same: in the network
dynamics the frozen nodes play a role only if the node `decides' to
copy a neighbor (probability $1-p$). Here mutation acts even if the
allele is passed from the focal individual $i$ to the offspring. The
master equation that includes mutation is therefore slightly different.
Using $p=1/2$, which is appropriate for unbiased reproduction, we have:

\begin{displaymath}
\begin{array}{ll}
P_{t+1}(m) &=  \displaystyle{P_t(m)\left\{
\frac{1}{2}\left(\frac{m}{N}\right)(1-\mu_+) +
\frac{1}{2}\left(\frac{N-m}{N}\right)(1-\mu_-) + \right.} \\ \\
& \displaystyle{ \frac{1}{2} \left(\frac{m}{N}\right) \left[
\left(\frac{m-1}{N-1}\right)(1-\mu_+) +
\left(\frac{N-m}{N-1}\right) \mu_- \right] + } \\ \\
& \displaystyle{ \left. \frac{1}{2} \left(\frac{N-m}{N}\right) \left[
\left(\frac{N-m-1}{N-1}\right)(1-\mu_-) +
\left(\frac{m}{N-1}\right) \mu_+ \right]\right\} +} \\ \\
& P_t(m-1) \displaystyle{ \left(\frac{N-m+1}{N}\right) \left[\frac{\mu_-}{2}
+\frac{1}{2} \left(\frac{m-1}{N-1}\right)(1-\mu_+) +
\frac{1}{2} \left(\frac{N-m}{N-1}\right)\mu_-\right] } \, + \\ \\
& P_t(m+1) \displaystyle{ \left(\frac{m+1}{N}\right) \left[\frac{\mu_+}{2}
+\frac{1}{2} \left(\frac{N-m-1}{N-1}\right)(1-\mu_-) +
\frac{1}{2} \left(\frac{m}{N-1}\right)\mu_+\right]} \;.
\end{array}
\end{displaymath}

The first terms can be understood as follows: if the population has $m$
individuals with allele $A_1$ at time $t$, it can remain that way in
the next time step in several ways. First, if $x_i=1$  (probability
$m/N$) the offspring can keep the allele $A_1$ if it gets it from
individual $i$ (probability $1/2$) and it does not mutate after
reproduction (probability $1-\mu_+$). Similarly, if $x_i=0$
(probability $(N-m)/N$) the offspring can keep the allele $A_0$ if it
gets it from individual $i$ (probability $1/2$) and does not mutate
after reproduction (probability $1-\mu_-$). The other terms have
similar interpretations.

This equation is greatly simplified when written in matrix form. We
obtain
\begin{equation}
P_{t+1} = U P_t \equiv \left[ 1 - \frac{(1+2\bar{\mu})}{2N(N-1)} A\right] P_t
\end{equation}
where the non-zero elements of $A$ are given by
\begin{displaymath}
\begin{array}{l}
A_{m,m} = 2m(N-m) +N_1(N-m)+N_0 m \\
A_{m,m+1} = -(m+1)(N-m-1+N_0) \\
A_{m,m-1} = -(N-m+1)(m-1+N_1)
\end{array}
\end{displaymath}
with
\begin{equation}
\begin{array}{ll}
N_1 &\equiv \displaystyle{ \frac{2\mu_-(N-1)}{1-2\bar{\mu}} } \\ \\
N_0 &\equiv \displaystyle{\frac{2\mu_+(N-1)}{1-2\bar{\mu}} } \\ \\
\end{array}
\end{equation}
and
\begin{equation}
\bar{\mu} =\displaystyle{ \frac{\mu_++\mu_-}{2}}.
\end{equation}

This is identical to the original matrix $A$ of the network dynamics!
Therefore, all the known solutions of the network problem can be
directly transferred to the genetic problem via the above relation
between the mutation rates $\mu_-$ and $\mu_+$ and the frozen nodes
$N_0$ and $N_1$. These solutions are described in the next section.

%%%%%%%%%%%%%%%%%%%%%%%%%%%%%%%%%%%%%%%%%%%%%%%%%%%%%%%%%
%%%%%%%%%%%%%%%%%%%%%%%%%%%%%%%%%%%%%%%%%%%%%%%%%%%%%%%%%
\section{Equilibrium distribution}
\label{equi}

The cases $N_0=0$ or $N_1=0$, corresponding to $\mu_+=0$ or $\mu_-=0$,
are trivial since all individuals in the population will eventually
become identical, with allele $A_0$ or $A_1$ respectively. If $N_0$ and
$N_1$ are both zero the individuals will also eventually become
identical, but the probability of each outcome, all $A_0$ or all $A_1$,
depend on the initial distribution of alleles in the population.

If $N_0$ and $N_1$ are both non-zero, the probability of finding $m$
nodes in state $1$, or $m$ individuals with allele $A_1$, in
equilibrium is given by \cite{aguiar2007,ewens,gill}
\begin{equation}
\rho(k) =  A(N,N_0,N_1) ~ \frac{\Gamma(N_1+k)~\Gamma(N + N_0 - k)}
{\Gamma(N-k+1)~\Gamma(k+1)}.
\label{probn}
\end{equation}
where
\begin{equation}
A(N,N_0,N_1) = \frac{\Gamma(N+1)~\Gamma(N_0+N_1)}{\Gamma(N + N_0 + N_1)
~\Gamma(N_1)~\Gamma(N_0)}.
\end{equation}
is a normalization constant and $\Gamma(x)$ is the Gamma function. This
result is valid even if $N_0$ and $N_1$ are not integers. In a real
network system, when $N_0$ and $N_1$ are integer numbers, the Gamma
functions can be replaced by factorials.

Notice that, because of the mutation rates (or frozen nodes), a
particular realization of the dynamics will never stabilize in any
state: the number of individuals with allele $A_1$ will always change.
The probability of finding the population with $m$ alleles $A_1$,
however, is independent of the time, and given by the expression above.
One interesting feature of this solution is that for $N_0=N_1=1$ we
obtain $\rho(m)=1/(N+1)$ for all values of $m$, meaning that all states
are equally likely, no matter how large is the population.

The mean value $m_0=\sum_m m \rho(m)$ and the variance $\sigma_2
=\sum_m m^2 \rho(m) - \bar{m}^2$ can also be calculated explicitly. We
obtain
\begin{equation}
m_0 = N \frac{N_1}{N_0+N_1}.
\label{avem}
\end{equation}
and
\begin{equation}
\displaystyle{\sigma_2 =
\frac{NN_1N_0(N_1+N_0+N)}{(N_1+N_0)^2(1+N_1+N_0)}}
\label{eq6a}
\end{equation}
Higher order correlations can also be calculated explicitly, but the
results become progressively more complicated.

Figures 1 and 2 show a few examples of the distribution $\rho(m)$ for a
network with $N=100$ and various values of $N_0$ and $N_1$.

\begin{figure}
   \includegraphics[clip=true,width=6cm,angle=-90]{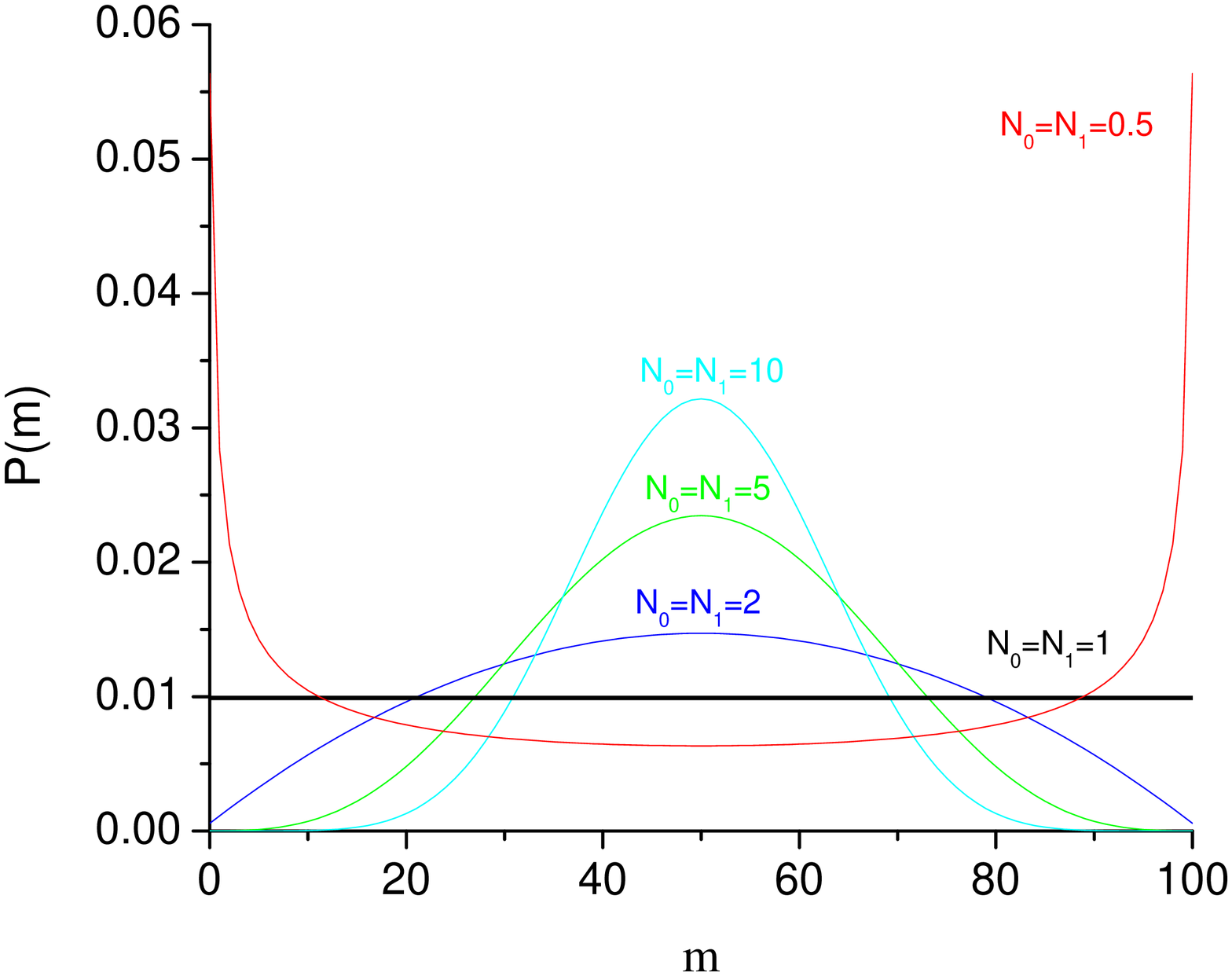}
   \includegraphics[clip=true,width=6cm,angle=-90]{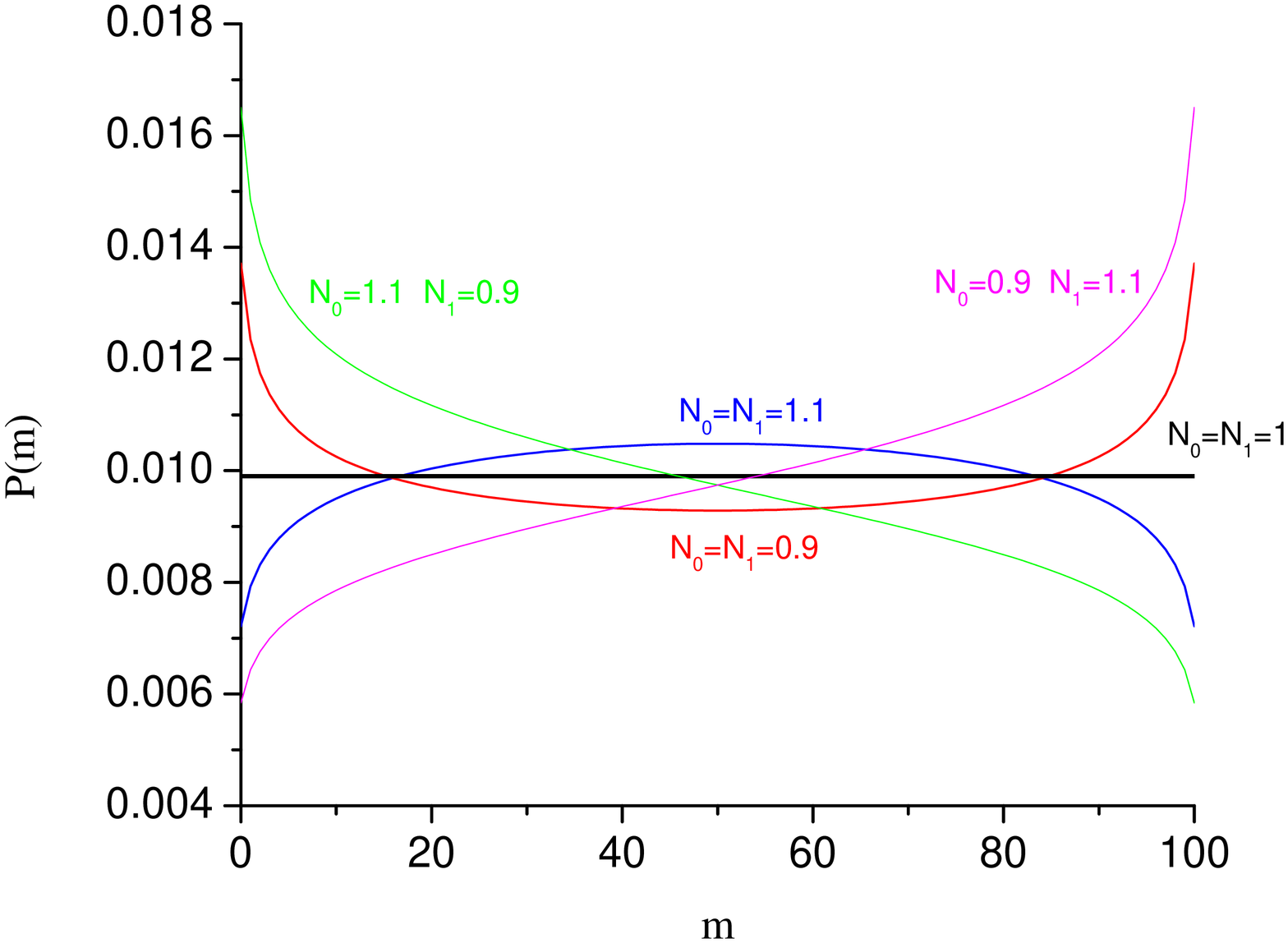}
   \caption{Asymptotic probability distribution for a network with
   $N=100$ nodes and several values of $N_0$ and $N_1$.}
\end{figure}

If $N$ is very large $\rho(m)$ peaks around $m_0$ and can be
approximated by a Gaussian:
\begin{displaymath}
\rho(m) =  \rho_0 ~ \exp{-\left[\frac{(m-m_0)^2}{2\Delta^2}\right]}.
\end{displaymath}
with
\begin{displaymath}
\Delta = \left[ \frac{N N_0 N_1 (N+N_0+N_1)}{(N_0+N_1)^3}
\right]^{1/2}
\end{displaymath}
and
\begin{displaymath}
\rho_0 = \frac{1}{\sqrt{2\pi}\Delta}.
\end{displaymath}
In terms of the continuous variables $x=m/N$, $n_0=N_0/N$ and
$n_1=N_1/N$ we can also write
\begin{displaymath}
\rho(x) =  \rho_0 ~ \exp{-\left[\frac{(x-x_0)^2}{2\delta^2}\right]}.
\end{displaymath}
with
\begin{displaymath}
\delta = \left[ \frac{n_0 n_1 (1+n_0+n_1)}{N(n_0+n_1)^3}
\right]^{1/2}
\label{delta}
\end{displaymath}
$x_0=m_0/N$ and $\rho_0 = 1/\sqrt{2\pi}\delta$, showing that the width
of the distribution goes to zero as $N$ goes to infinity, in agreement
with the Hardy-Weinberg law.

%%%%%%%%%%%%%%%%%%%%%%%%%%%%%%%%%%%%%%%%%%%%%%%%%%%%%%%%%
%%%%%%%%%%%%%%%%%%%%%%%%%%%%%%%%%%%%%%%%%%%%%%%%%%%%%%%%%
\section{Structured networks}
\label{structured}

For networks that are not fully connected the effect of the frozen
nodes is amplified. To see this we note that the probability that a
free node copies a frozen node is $P_i=(N_{0}+N_{1})/(N_{0}+N_{1}+k_i)$
where $k_i$ is the degree of the node. For fully connected networks
$k_i=N-1$ and $P_i\equiv P_{FC}$. For general networks an average value
$P_{av}$ can be calculated by replacing $k_i$ by the average degree
$k_{av}$. We can then define effective numbers of frozen nodes,
$N_{0ef}$ and $N_{1ef}$, as being the values of $N_{0}$ and $N_{1}$ in
$P_{FC}$ for which $P_{av} \equiv P_{FC}$. This leads to
\begin{equation}
N_{0ef} = f N_{0}, \qquad \qquad N_{1ef} = f N_{1}
\label{eq:rescaling}
\end{equation}
where $f=(N-1)/k_{av}$. Corrections involving higher moments can be
obtained by integrating $P_i$ with the degree distribution and
expanding around $k_{av}$.
\begin{figure}
   \includegraphics[clip=true,width=4cm,angle=-90]{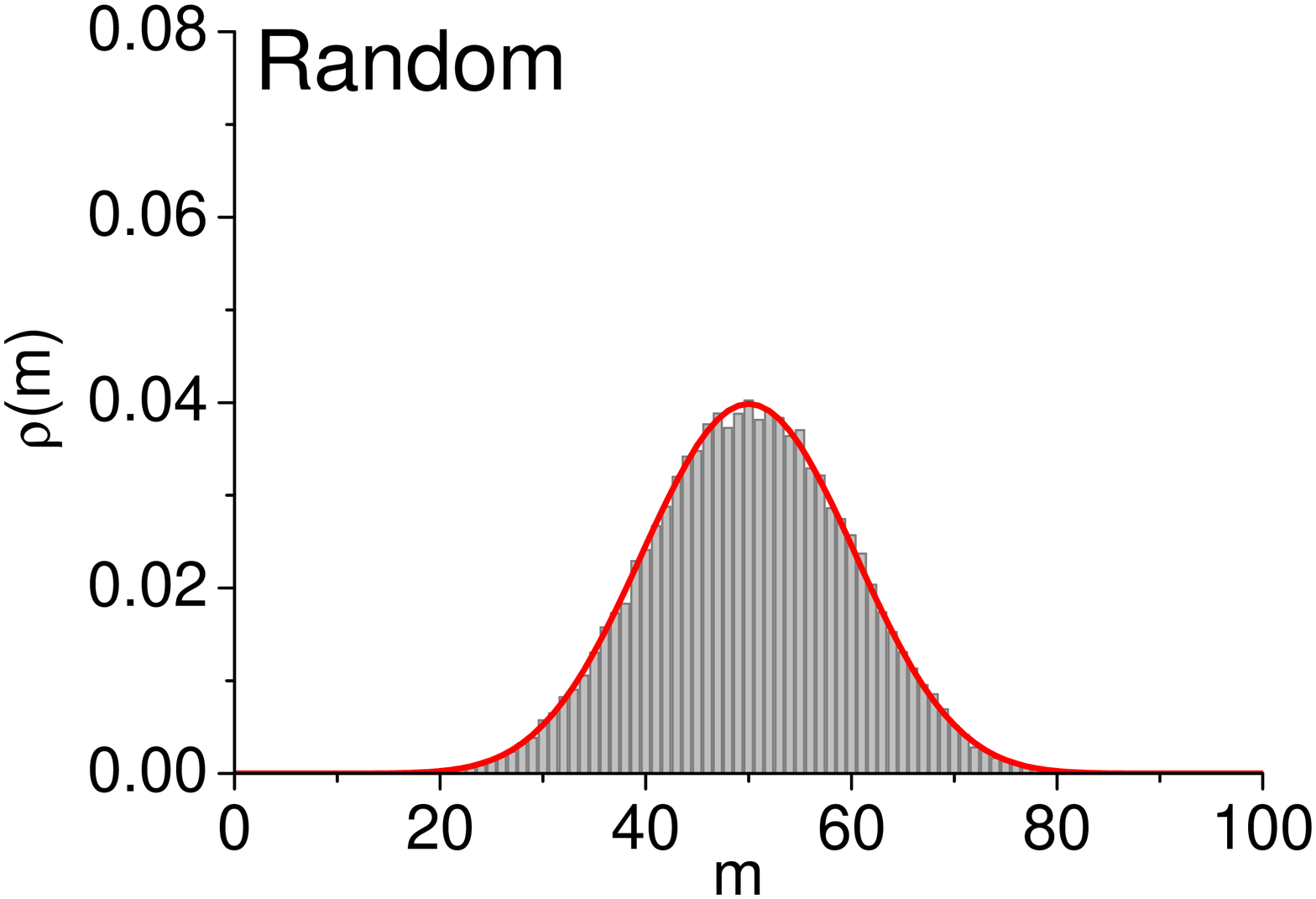}
   \includegraphics[clip=true,width=4cm,angle=-90]{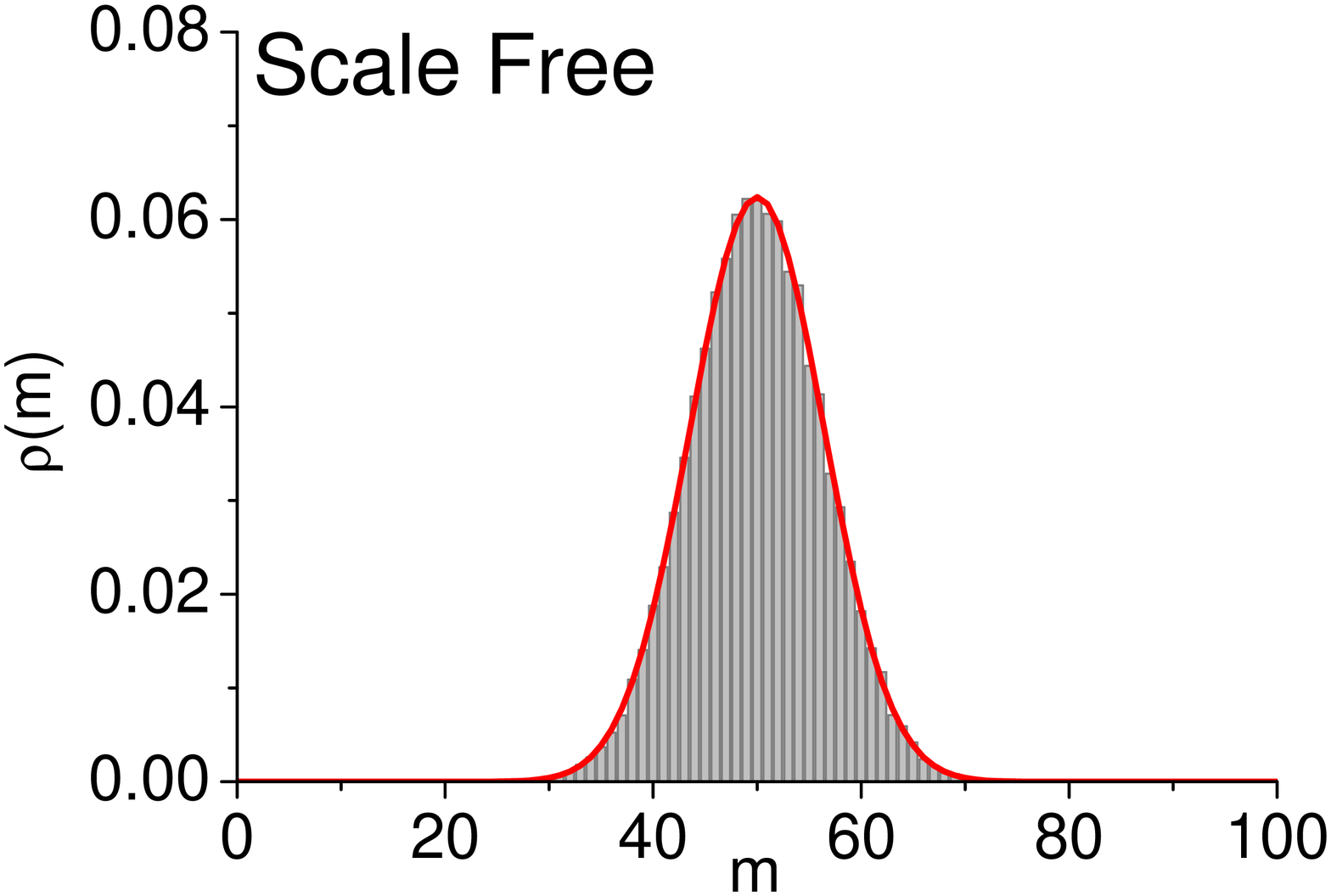}
   \includegraphics[clip=true,width=4cm,angle=-90]{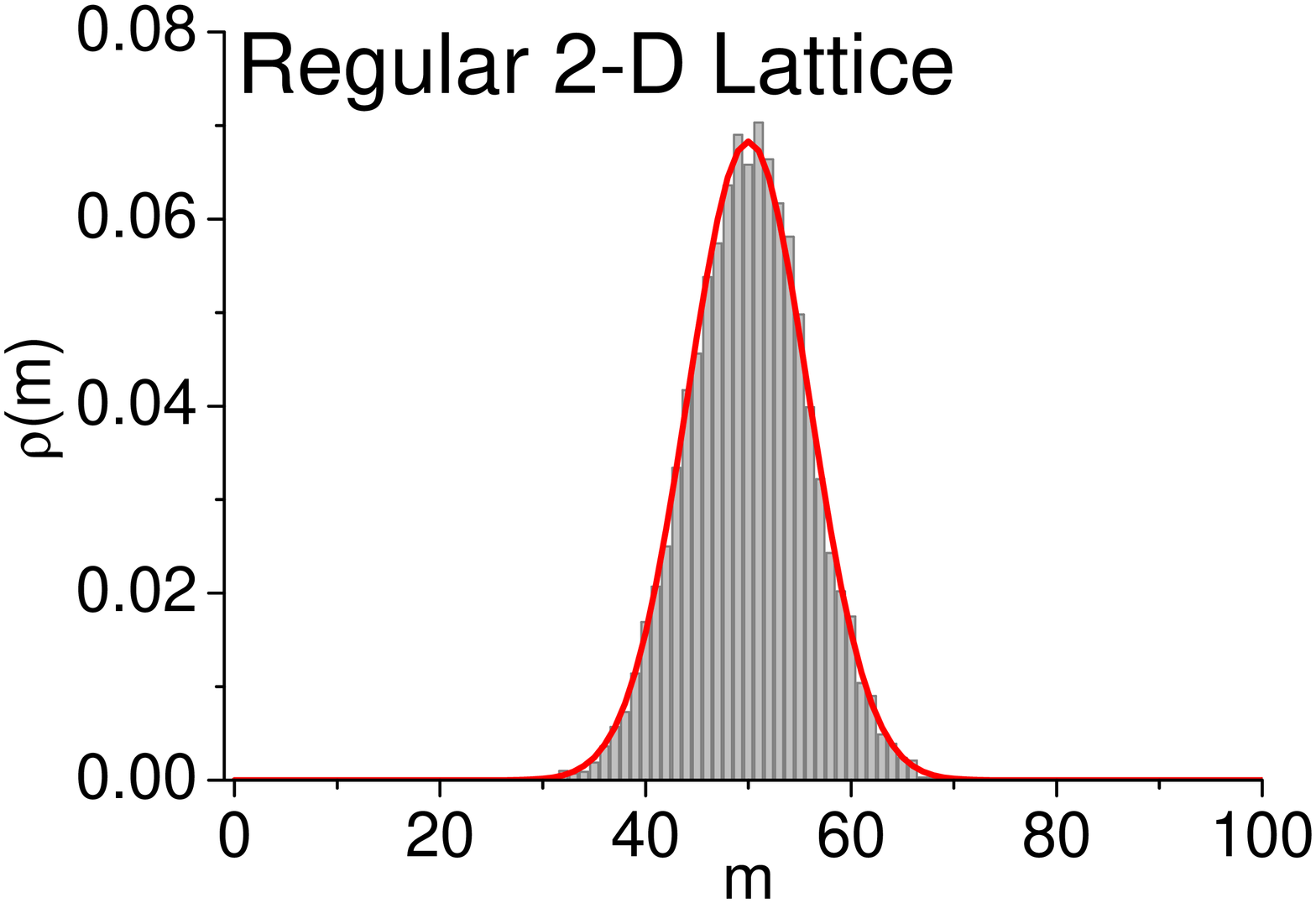}
   \includegraphics[clip=true,width=4cm,angle=-90]{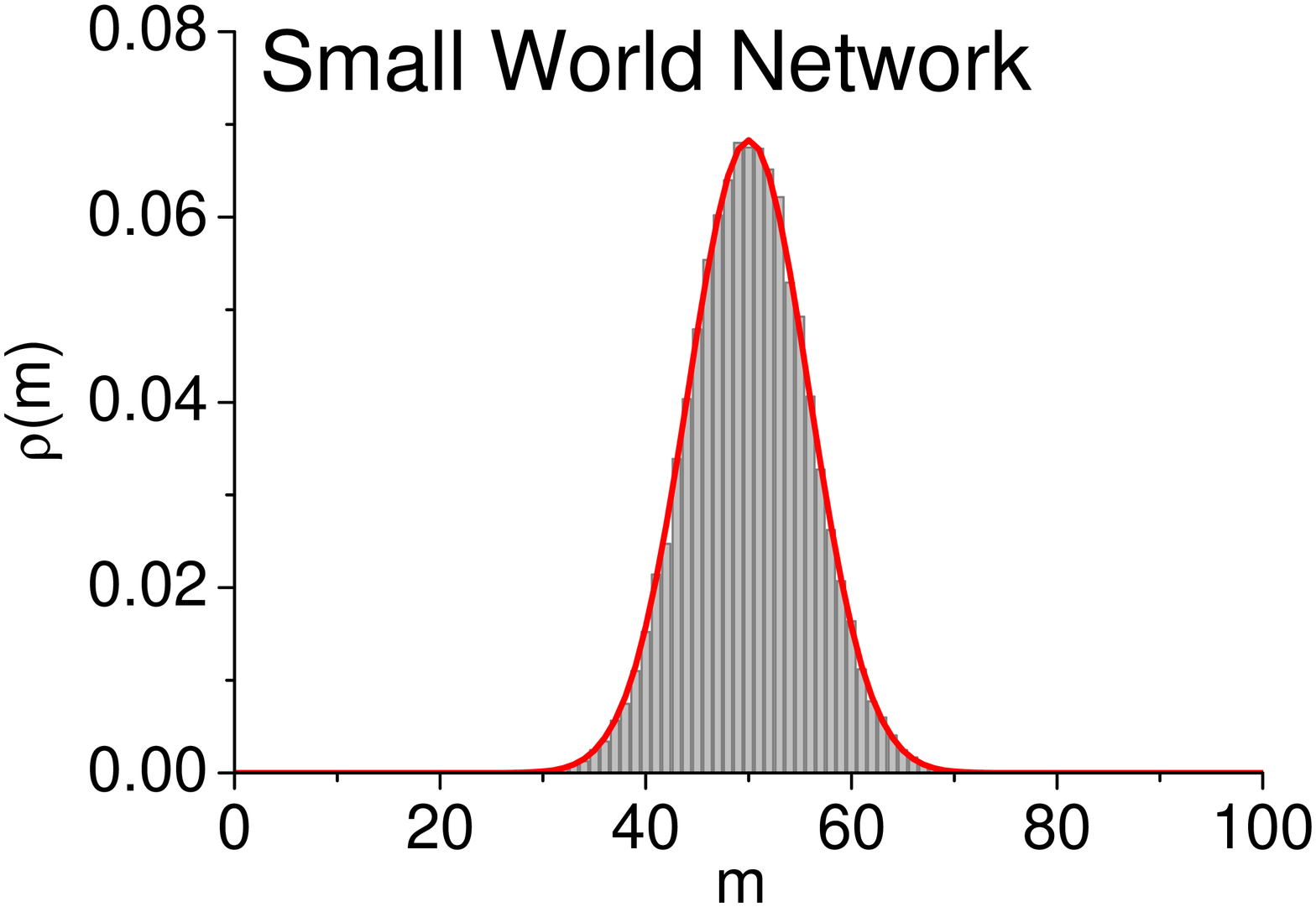}
   \caption{Equilibrium probability distribution for networks with
   different topologies. In all cases $N=100$, $N_0=N_1=5$, $t=10,000$,
   and the number of simulations is $50,000$. The theoretical (red) curve
   is drawn with effective numbers of frozen nodes $N_{0ef}=f N_0$ and
   $N_{1ef}=f N_1$: (a) random network $N_{0ef}=N_{1ef}=17$; (b)
   scale-free $N_{0ef}=N_{1ef}=82$; (c) regular 2-D lattice
   $N_{0ef}=N_{1ef}=140$; (d) small world network $N_{0ef}=N_{1ef}=140$.}
   \label{fig2}
\end{figure}

Figure \ref{fig2} shows examples of the equilibrium distribution for
four different networks with $N=100$ and $N_0=N_1=5$. Panel (a) shows the
result for a random network constructed by connecting any pair of nodes
with probability $0.3$. In this case $k_{av}=29.7$ and $f=3.3$. The
theoretical result was obtained with Eq.~(\ref{probn})
with $N_{0ef}=N_{1ef}=17$. For a scale-free network (panel (b)) grown
from an initial cluster of 6 nodes adding nodes with 3 connections each
following the preferential attachment rule \cite{bararev}, $f=99/6$ and
the effective values of $N_0$ and $N_1$ are approximately 82. Panel (c)
shows the probability distribution for a 2-D regular lattice with $10
\times 10$ nodes connected to nearest neighbors for which $k_av=3.6$ (the
nodes near the border have less than 4 links) $f=99/3.6\approx 28$.
Finally, panel (d) shows a small world version of the regular lattice
\cite{bararev}, where 30 connections were randomly re-allocated, creating
shortcuts between otherwise distant nodes. These results show that the
approximate re-scaling of frozen nodes (or, equivalently, the mutation
rates) is accurate for many network topologies. Still, extreme
cases such as a star network do present different distributions and
this is confirmed by simulations.

%%%%%%%%%%%%%%%%%%%%%%%%%%%%%%%%%%%%%%%%%%%%%%%%%%%%%%%%%
%%%%%%%%%%%%%%%%%%%%%%%%%%%%%%%%%%%%%%%%%%%%%%%%%%%%%%%%%
\section{Speciation and biodiversity}
\label{speciation}

In the last sections we derived two important theoretical results: (a) the
connection between the process of influence dynamics on networks and the
Moran model; (b) the approximate equilibrium distribution for structured
networks, obtained by re-scaling the number of frozen nodes. We
will show now that these two results allow us to infer important
properties about the genetic evolution of spatially extended populations.

It has been recently shown \cite{doursat2008,aguiar2009} that when mating
is constrained by both spatial and genetic proximity between individuals,
neutral evolution by drift alone might lead to speciation, i.e., to the
spontaneous break up of the population into reproductively isolated
clusters. Moreover, the patterns of abundance distributions generated by
this mechanism are compatible with those observed in nature
\cite{aguiar2009}.

Neutral theories of biodiversity have become rather sophisticated
\cite{hubbell2001}, heating the neutralist-selectionist debate
\cite{gavrilets2000,banavar2009,kopp2010,terSteege2010,etienne2010}. In
what follows we discuss the process of neutral speciation promoted by
spatial and genetic constraints, termed {\it topopatric speciation}, in
the light of the theory developed above.

To make the analysis simpler we will restrict ourselves to the case of
symmetric mutation rates, $\mu_-=\mu_+ \equiv \mu$ or, equivalently,
equal number of frozen nodes $N_0=N_1 \equiv N_z$. In this case the
connection between mutations and frozen nodes simplifies to
\begin{equation}
N_z = \displaystyle{\frac{2\mu(N-1)}{1-2\mu} }.
\label{nmu}
\end{equation}

Let $P_{id}$ be the probability that two individuals picked at random in
the population have identical genes at equilibrium. This is given by
the sum of the probabilities that their alleles are are both $A_1$ or both
$A_0$:
\begin{displaymath}
\begin{array}{ll}
P_{id} &= \sum_{m=0}^N \rho(m)\left[\frac{m}{N}\frac{m-1}{N-1}+
\frac{N-m}{N}\frac{N-m-1}{N-1} \right] \\  \\
&= 1 + \frac{2}{N(N-1)}\left[\sigma^2+\langle m \rangle^2-
N\langle m \rangle\right].
\end{array}
\label{eq1}
\end{displaymath}
Using equations (\ref{avem}), (\ref{eq6a}) and (\ref{nmu}) we obtain
\begin{equation}
P_{id} = \frac{1+N_z}{1+2 N_z} = \frac{1+2\mu(N-2)}{1+2\mu(2N-3)}.
\end{equation}
The probability that the two individuals are different, which is the
heterozigosity, is
\begin{equation}
P_{ht} = 1 - P_{id} =
\frac{2\mu(N-1)}{1+2\mu(2N-3)} \approx \frac{2\mu N}{1+4\mu N}
\end{equation}
where the approximation holds for $N>>1$.

Consider now a population in equilibrium where the $N$ individuals have
$B$ independent genes
\cite{higgs1991,zhang1997,jensen2002,jain2007,doursat2008,aguiar2009}.
The average genetic distance between two individuals is
\begin{equation}
\langle d\, \rangle = B P_{ht} \approx
\frac{B}{2} \left(\frac{4\mu N}{1+4\mu N}\right).
\label{eq5a}
\end{equation}

This expression provides a connection between the size of the population
and the average genetic distance between individuals, which is a
measure of diversity within the population. Two interesting relations
can be derived from this equation: first, for given $B$ and $\mu$ we
can calculate the size $N_G$ that corresponds to a particular
average genetic distance $\langle d\, \rangle = G$:
\begin{equation}
N_G = \frac{G}{2\mu(B-2G)}.
\label{eq7}
\end{equation}
Second, for given $N$ and $B$ we calculate the mutation rate
$\mu_G$ that corresponds to $\langle d\, \rangle =
G$:
\begin{equation}
\mu_G = \frac{G}{2N(B-2G)}.
\label{eq7a}
\end{equation}
Notice that $N_G \mu = N \mu_G$.

When mating in panmictic populations is constrained by genetic proximity
between individuals, so that pairs whose genetic distance is larger than
$G$ are incompatible, the distribution of genetic distances stays very
close to $\langle d\, \rangle = G$, as if the genome had an effective
size $B_{ef}=2G$.  On the other hand, if mating is
constrained by spatial proximity, the effective mutation rate tends to
increase. Indeed, spatial restriction in mating corresponds to influence
processes on networks constructed over regular lattices, which amplifies
the effect of frozen nodes and, therefore, of mutations.

Consider a square lattice with $L^2$ nodes and periodic boundary
conditions where each node is connected only to neighbors which are
within a distance $S$ from itself (measured in units of lattice
spacing). Let $N$ be the number of individuals in the population, so that
the density is $\rho = N/L^2$. The area where an individual can
look for a mate, its `mating neighborhood', is approximately $\pi S^2$,
which is also the average degree $k_{av}$ of the network.

According to our discussion in section \ref{structured}, this can be
modeled as fully connected network with effective number of frozen
nodes
\begin{equation}
N_{ef} = f N_z = \frac{N-1}{k_{av}} \, N_z
\approx \frac{N }{\pi S^2}N_z.
\label{eq8}
\end{equation}
The corresponding effective mutation rate is obtained from (\ref{nmu})
\begin{displaymath}
N_{ef} = \frac{2\mu_{ef}(N-1)}{1-2\mu_{ef}}
\end{displaymath}
which gives
\begin{equation}
 \mu_{ef} = \frac{f}{1+2\mu(f-1)} \mu \approx \frac{\mu f}{1+2\mu f}.
\label{muef}
\end{equation}
Note that $\mu_{ef} \rightarrow 1/2$ if $\mu f >> 1$.

When mating between individuals is constrained by their spatial distance,
as measured by the parameter $S$, the effective mutation rate
(\ref{muef}) can be dramatically enhanced with respect to a panmictic
population. This, in turn, increases the average genetic distance between
individuals, which approaches $B/2$ for large populations and fixed
$k_{av}$ (corresponding to large values of $N_z$). The distribution of
genetic distances approaches a broad symmetric distribution.

On the other hand, if mating is constrained only by the genetic distance
between individuals, the distribution of genetic distances shrinks to
about $G$. This corresponds to an effective shrink in genome size from $B$
to $2G$.

When both spatial and genetic restrictions are present, as in
\cite{aguiar2009}, the population feels a large effective mutation rate,
tending to spread out the genome distribution. On the other hand, the
individuals are compelled by the mating condition to stay genetically
close to each other. The only stable outcome of these opposing forces is
the formation of local groups where  $\langle d \, \rangle \leq G$ within
the group but $\langle d \, \rangle > G$ among groups. This characterizes
the groups as reproductively isolated from each other and, therefore, as
separate species.

The average number of individuals in each group is given approximately by
$N_G$ (\ref{eq7}), which is usually much smaller than $N$. This also
implies that the individuals within groups are highly connected to each
other, so that $f \approx 1$ and $\mu_{ef} \approx \mu$, restoring the
equilibrium of the system.

The conditions for speciation can be estimated as follows. When $S$ is
very large, the effect of the genetic mating restriction is to reduce
the effective size of the genome, $B_{ef}$, from $B$ to $2G$, so that,
from equation (\ref{eq5a}), $\langle d \, \rangle$ is at most $G$. As
$S$ is reduced, the effective mutation rate increases and new genes are
incorporated into the effective genome, increasing the average genetic
distance between individuals. When $\langle d \, \rangle$ becomes
larger than about $2G$ the population can no longer hold itself
together and splits. This has been confirmed by numerical simulations.
We write
\begin{equation}
B_{ef} = 2G + (B-2G) {\cal P}
\label{eqbef}
\end{equation}
where ${\cal P}$ is the probability that a new gene is fixed into the
effective genome.

${\cal P}$ goes to zero for large values of $S$ and reaches one for
small $S$. It must depend only on the mutation rate $\mu$, genome
length $B$ and the size of the local mating population $N_S \equiv \pi
S^2 \rho = \pi S^2 N / L^2$. This local mating population has to be at
least 2, otherwise mating is not possible. More generally, if the
minimum number of potential mates for reproduction is $P$ we can define
the minimum $S$ by $\pi S_{min}^2 \rho = P$, or
\begin{equation}
 S_{min} = L \sqrt{P/\pi N}.
\end{equation}

${\cal P}$  must be small if the local mating population is large. On
the other hand, it must increase with the mutation rate and size of the
genome. We may therefore write the ansatz
\begin{displaymath}
{\cal P} = \exp{ \left\{-c \left[\frac{\pi (S-S_{min})^2 N/L^2)}{B \mu}
\right]^2 \right\} }
\end{displaymath}
or
\begin{equation}
{\cal P} = \exp{ \left\{-\frac{\pi^2 (S-S_{min})^4 N^2}{\gamma^4 L^4
B^2 \mu^2} \right\} }.
\label{eqpp}
\end{equation}
where the constant of proportionality $c$ is rewritten by as
$\gamma^{-4}$ for convenience. The exponential dependence of ${\cal P}$
on the square of $N_S/B\mu$ is suggested by numerical simulations.
\begin{figure}
   \includegraphics[clip=true,width=12cm]{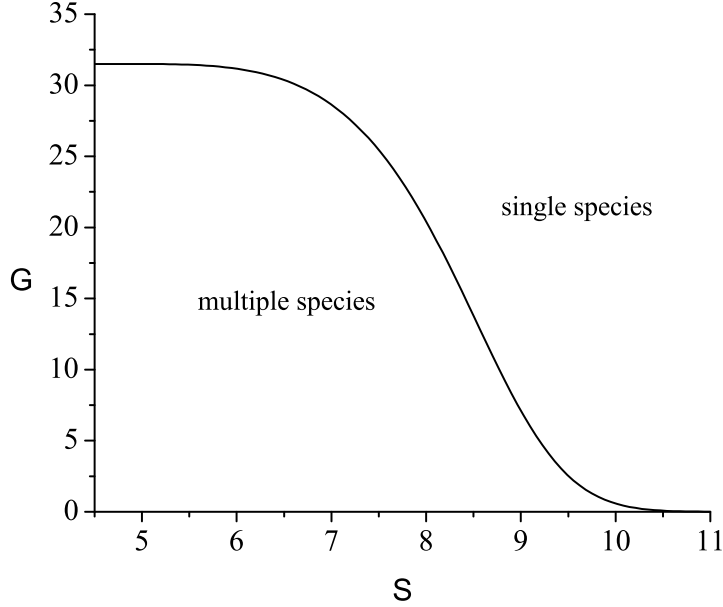}
   \caption{Parameter region where speciation is possible according with
equation (\ref{gs}). In this example $N=2000$, $\mu=0.001$, $B=125$,
$L=128$ and $P=8$ ($S_{min}=4.6$) and $\gamma=6.6$ (see \cite{aguiar2009}).}
   \label{fig3}
\end{figure}

The condition for speciation is
\begin{displaymath}
\langle d \,\rangle = \frac{B_{ef}}{2} \left(\frac{4\mu_{ef}
N}{1+4\mu_{ef} N}\right) \gtrsim 2G.
\end{displaymath}
Since the $\mu N$ is usually of order 1 in most simulations, and
$\mu_{ef} >> \mu$, the factor $4\mu_{ef} N/(1+4\mu_{ef} N)$ can be
safely approximated by 1. Using equations (\ref{eqbef}) and (\ref{eqpp})
we obtain
\begin{displaymath}
\frac{\pi^2 (S-S_{min})^4 N^2}{\gamma^4 L^4 \mu^2 B^2} \lesssim
\log{\left(\frac{B-2G}{2G}\right)}
\end{displaymath}
or
\begin{equation}
S \lesssim S_{min} + \gamma L \sqrt{\frac{B \mu}{N \pi}}\left[
\log{\left(\frac{B-2G}{2G}\right)} \right]^{1/4} \equiv S_{c}(G).
\label{sg}
\end{equation}
Inverting this equation we obtain
\begin{equation}
G \lesssim \frac{B/2}{1+
\exp{ \left( \frac{\pi^2 N^2 (S-S_{min})^4}{\gamma^4 \mu^2 B^2 L^4}
\right)}} \equiv G_{c}(S)
\label{gs}
\end{equation}
which gives the minimum value of $G$ for a given $S$.

Equation (\ref{sg}) gives the maximum size of the mating neighborhood
for which speciation is possible. This analytical result describes the
dependence of speciation on 6 model parameters: $B$, $G$, $\mu$, $P$,
$L$ and $N$. It provides a very good quantitative estimate for the
parameter region where speciation is possible, as illustrated in figure
\ref{fig3}. The result also incorporates cutoffs at $G = B/4$ and at
$S_{min}$, which are in agreement with numerical simulations
\cite{aguiar2009}. Furthermore it also gives the scaling dependence of
$S_{c}$ on these various parameters. In particular, it predicts
speciation at large values of $S$ if $B$ is sufficiently large. This
corroborates the results in \cite{higgs1991,higgs1992} but shows that
such space-independent speciation occurs only for very large values of
$B$, since $S$ increases with $B^{1/2}$.

Our analytical result constitute an important addition to the
simulations presented in \cite{aguiar2009} and contribute to the
understanding of the significant role of drift in speciation
\cite{coyne,aguiar2009,doursat2008,hubbell2001,higgs1991,kopp2010,terSteege2010}.
Equation (\ref{sg}) identifies the combination of parameters that makes
this possible. For example, low mutation rates, that hinder speciation,
can be compensated by a large number of participating genes or by low
population density.

%%%%%%%%%%%%%%%%%%%%%%%%%%%%%%%%%%%%%%%%%%%%%%%%%%%%%%%%%
%%%%%%%%%%%%%%%%%%%%%%%%%%%%%%%%%%%%%%%%%%%%%%%%%%%%%%%%%
\section{Conclusions}
\label{conc}

The process of speciation underlies the creation of the tree of life.
Fossil records and molecular analysis allow the construction of
detailed phylogenetic trees linking species to their ancestors,
identifying the branching points of speciation. The way speciation
occurred in each case, however, is rarely known with certainty and
several mechanisms have been considered. A recently proposed mechanism
of speciation \citep{aguiar2009} demonstrated that a spatially extended
population can break up spontaneously into species when subjected to
mutations and to spatial and genetic mating restrictions, even in the
absence of natural selection. Numerical simulations have shown that
this mechanism, termed topopatry, occurs for a restricted range of
parameters, that include population size $N$, mutation rate $\mu$ and
the parameters $S$ and $G$ controlling the spatial and genetic mating
restrictions.

In this paper we have introduced a mapping of genetic dynamics in an
evolving population onto the dynamics of influence on a network, and
used this mapping to analytically study the process of topopatric
speciation. This mapping gives, to our knowledge, the first complete
solution of the Moran model, providing an elegant representation of the
complete set of eigenvectors of the problem.

We have shown that, while fully connected networks correspond to
panmictic populations, certain structured networks can be mapped into
dynamic spatially extended populations. Moreover, the mapping shows
that limiting mating to a fraction of the total population by network
connections increases the effective mutation rate as compared to the
panmictic case, and increases the genetic diversity of the population.
By extending the model from one to multiple independent biallelic
genes, we have shown that a genetic restriction on mating decreases the
effective size of the genome, decreasing diversity. These opposing
forces are resolved not by compromise but by pattern formation,
breaking up the population into multiple species. This process, and its
dependence on the most relevant characteristics of the population, is
accurately described by equation (\ref{gs}). This equation provides a new
and important tool to understand neutral speciation, revealing explicitly
the relationships among the parameters involved in the process, and the
interplay of genetic processes whose opposition leads to spontaneous
speciation. \\ \\

\noindent ACKNOWLEDGMENTS

\noindent It is a pleasure to thank Elizabeth M. Baptestini for helpful
comments. MAMA acknowledges financial support from CNPq and FAPESP.

%%%%%%%%%%%%%%%%%%%%%%%%%%%%%%%%%%%%%%%%%%%%%%%%%%%%%%%%%%%%%%%%

\end{document}